\documentclass[epj]{webofc}
\usepackage[varg]{txfonts}   % Web of Conferences font
\usepackage{graphicx}

\begin{document}

\title{Neutrino Sources from a Multi-Messenger Perspective}

\author{Markus Ahlers\thanks{\email{markus.ahlers@nbi.ku.dk}}}

\institute{Niels Bohr International Academy \& Discovery Centre, Niels Bohr Institute,\\University of Copenhagen, Blegdamsvej 17, DK-2100 Copenhagen, Denmark}

\abstract{The field of high-energy neutrino astronomy is undergoing a rapid evolution. After the discovery of a diffuse flux of astrophysical TeV-PeV neutrinos in 2013, the IceCube observatory has recently found first compelling evidence for neutrino emission from blazars. In this brief review, I will summarize the status of these neutrino observations and highlight the strong role of multi-messenger astronomy for their interpretation.}

\maketitle

\section{Introduction}\label{intro}

Neutrinos are unique cosmic messengers. Their weak interaction with matter allows to probe the interior of dense sources, which are otherwise opaque to photons. Unlike $\gamma$-rays, which can suffer from strong absorption in the cosmic microwave background above TeV energies, high-energy neutrinos arrive at Earth from distant and therefore early sources of the Universe. And unlike cosmic rays, which are observed at Earth as a diffuse flux after repeated scattering in Galactic or extra-galactic magnetic fields, neutrinos point back to their sources. 

High-energy neutrino emission is expected from the interaction of cosmic ray (CR) nucleons with gas or radiation. These interactions produce pions, that subsequently decay via $\pi^+\to\mu^++\nu_\mu$ followed by $\mu^+\to e^++\nu_e+\bar\nu_\mu$ and the charge-conjugated processes. Typically, a neutrino from this production channel receives about 5\% of the energy from the initial CR nucleon. At the same time, CR interaction also produce neutral pions that decay as $\pi^0\to\gamma+\gamma$, where the individual $\gamma$-ray have an average energy fraction of 10\% with respect to the nucleon. Neutrinos are therefore tightly connected to other cosmic messengers and their sources need to be understood in the context of multi-messenger observations.

Presently, the most sensitive neutrino telescope in the TeV-PeV energy range is the IceCube Observatory located at the South Pole. The detector consists of 5,160 digital optical modules (DOMs) that fill a volume of about one cubic-kilometer of clear glacial ice. The DOMs are organized along 86 read-out and support cables (``strings'') at a depth between 1,450 and 2,450 meters below the surface. Most strings follow a triangular grid with a width of 125 meters, evenly spaced over the volume. The IceCube detector is fully instrumented since 2011, following a seven-year construction phase with incrementally increasing number of strings.

High-energy neutrinos are detected in IceCube via the Cherenkov light emission of charged particles produced by neutrino interactions in the vicinity of the detector. The most valuable events for neutrino astronomy are charged current interactions of muon neutrinos that can produce high-energy muons traversing the detector. These events allow for a good angular resolution of $\lesssim 0.4^\circ$ at the highest energy. Charged current interactions of tau neutrinos can also be visible by muon tracks in 18\% of the cases. All other neutrino interactions can be visible via the charged particles created in electro-magnetic cascades of the secondary electrons or in hadronic cascades from the breakup of the struck nucleons or the hadronic decay of taus. Whereas these cascade events have a poorer angular resolution in IceCube, at about $10^\circ-15^\circ$, they allow for a good estimator of the initial neutrino energy with a resolution of better than 15\,\%.

Cosmic ray interactions in the atmosphere produce a large background of muons and neutrinos that require a careful selection of astrophysical neutrino candidates. One possibility to eliminate the background of atmospheric muons is the selection of up-going muon track events. The remaining background of atmospheric neutrinos follows a steep spectrum that can be distinguished form a hard astrophysical neutrino flux via a statistical analysis. Another possibility consists of the selection of high-energy starting events (HESE). These events are required to have an interaction vertex inside a virtual outer veto layer of optical modules. The advantage of this selection is that atmospheric neutrinos from above the detector (Southern Hemisphere) are accompanied by atmospheric muons that trigger the veto~\cite{Schonert:2008is,Gaisser:2014bja}.

In the following we will summarize the status of diffuse neutrino observations with IceCube and recent evidence of neutrino emission from blazars.

\section{Diffuse Astrophysical Neutrinos}

\begin{figure}[t]\centering
\includegraphics[width=0.95\linewidth]{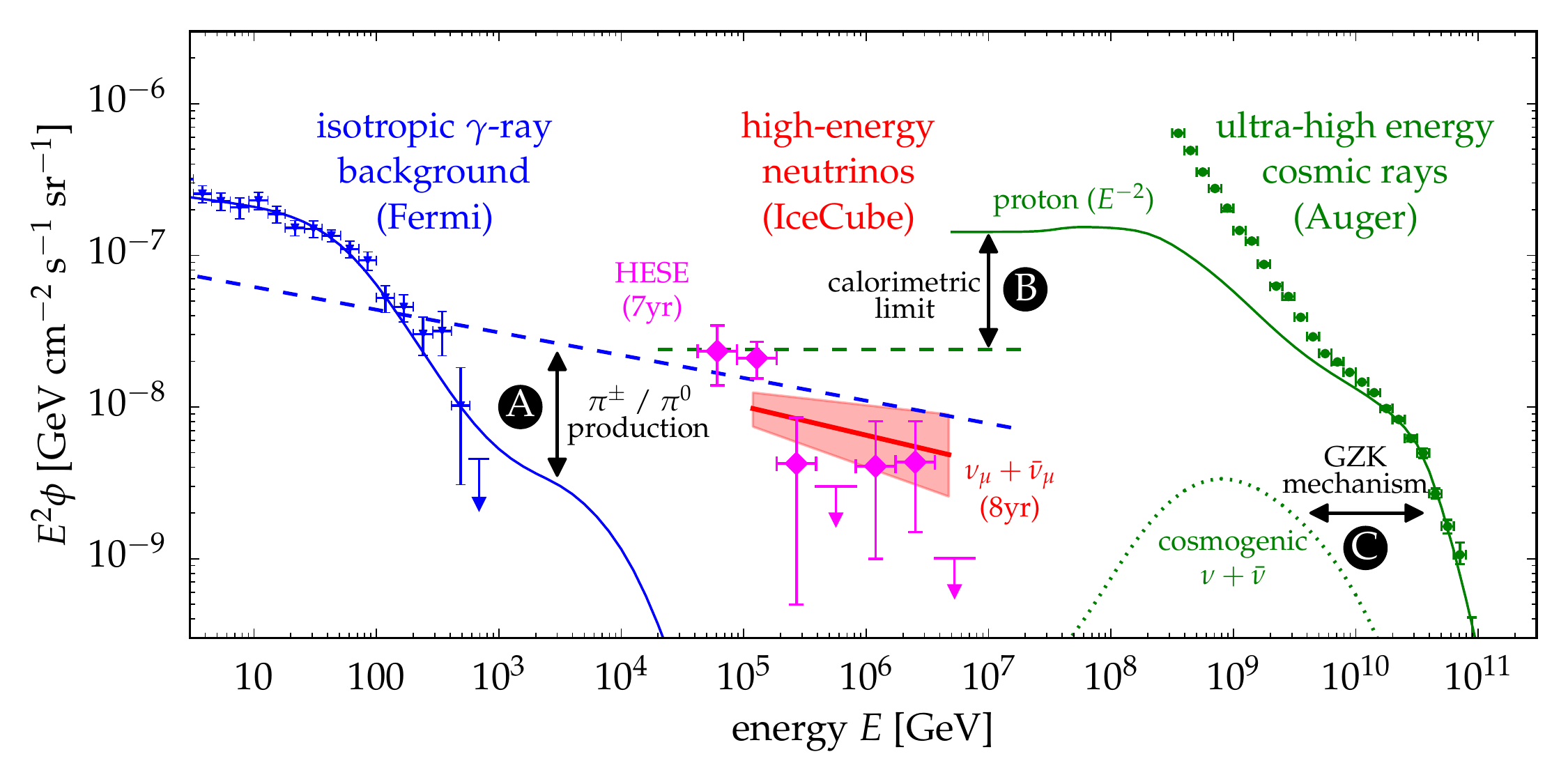}
\caption[]{The spectral flux ($\phi$) of neutrinos inferred from the eight-year upgoing track analysis (red fit) and preliminary results of the seven-year HESE analysis~\cite{Aartsen:2017mau} (magenta data) compared to the flux of unresolved extragalactic $\gamma$-ray sources~\cite{Ackermann:2014usa} (blue data) and ultra-high-energy cosmic rays~\cite{Aab:2015bza} (green data). The $\nu_\mu+\bar\nu_\mu$ spectrum is indicated by the best-fit power-law (solid line) and $1\sigma$ uncertainty range (shaded range). We highlight the various multimessenger relations: {\bf A:} The joined {production of charged pions ($\pi^\pm$) and neutral pions ($\pi^0$)} in cosmic-ray interactions leads to the emission of neutrinos (dashed blue) and $\gamma$-rays (solid blue), respectively. {\bf B:} Cosmic ray emission models (solid green) of the most energetic cosmic rays imply a maximal flux ({calorimetric limit}) of neutrinos from the same sources (green dashed). {\bf C:} The same cosmic ray model predicts the emission of cosmogenic neutrinos from the collision with cosmic background photons ({GZK mechanism}).}\label{fig1}
\end{figure}

In 2013, the IceCube Observatory reported first evidence of TeV-PeV astrophysical neutrinos in an analysis of two years of data based on the HESE selection~\cite{Aartsen:2013bka,Aartsen:2013jdh}. First evidence of this astrophysical flux in the up-going $\nu_\mu+\bar\nu_\mu$ sample was found shortly afterwards~\cite{Aartsen:2015rwa}. With continued observation, both analyses have now reached a significance of more than $5\sigma$~\cite{Aartsen:2014gkd,Aartsen:2016xlq,Aartsen:2017mau}. The diffuse flux inferred from the two event selections is shown in Fig.~\ref{fig1}. The HESE selection allows to study the astrophysical neutrino emission at lower energies due to the self-veto of atmospheric backgrounds. Above 100~TeV both results are consistent with a hard power-law spectrum, whereas at lower energies the HESE selection indicates a rising flux that deviates from a na\:ive power-law behavior.

Both analyses are consistent with a power-law extrapolation of the neutrino emission above 1~PeV. Neutrinos at these energies are associated with CR nuclei with an energy of the order of 20~PeV in the case of protons or 1~EeV in the case of iron. These CRs are therefore located in the CR spectrum between the ``knee'' ($2-5$~PeV) and the ``ankle'' ($3-5$~EeV), which is the transition region between the dominance of Galactic and extragalactic sources. Both types of sources have been considered in the literature as possible neutrino production sites and a summary can be found in the recent review~\cite{Ahlers:2018fkn}. However, the arrival directions of neutrinos are consistent with an isotropic distribution, after correcting for the angular acceptance of the detector and Earth absorption. This suggests that the signal is likely to originate from a population of relatively weak extragalactic sources. 

Figure~\ref{fig1} shows the diffuse neutrino flux $\phi$ in terms of the product $E^2\phi$, which is a measure of its local energy density. The energy density in cosmic neutrinos is comparable to that of the isotropic $\gamma$-ray background (IGRB) observed with the Fermi satellite~\cite{Ackermann:2014usa} (blue data in Fig.~\ref{fig1}) and to that of ultra-high-energy (UHE) CRs (above $10^{9}$~GeV) observed, {\it e.g.}, by the Auger observatory~\cite{Aab:2015bza} (green data in Fig.~\ref{fig1}). We will highlight in the following that this might indicate a common origin of these signals, which provides excellent conditions for multi-messenger studies.
 
The simultaneous production of neutral and charged pions in CR interactions suggests that the sources of high-energy neutrinos could also be strong TeV-PeV $\gamma$-ray emitters. For extragalactic scenarios, this $\gamma$-ray emission is not directly observable because of the strong absorption of photons by $e^+e^-$ pair production in the extragalactic background light (EBL) and the cosmic microwave background. The high-energy leptons initiate electromagnetic cascades of repeated inverse-Compton scattering and pair production in the CMB that eventually yield photons that contribute to the Fermi $\gamma$-ray observations in the GeV-TeV range. 

The corresponding $\gamma$-ray and neutrino fluxes of a simple power-law emission model are shown as solid and dashed blue lines, respectively, in Fig.~\ref{fig1}. The $\gamma$-ray flux is normalized to the Fermi data to indicate the maximal contribution of neutrinos. The observed neutrino flux above 100~TeV is very close to the corresponding upper limit (dashed blue line), which would imply a large contribution of the underlying source population to the $\gamma$-ray background. The neutrino flux below $100$~TeV slightly overshoots this bound and poses a challenge for this type of emission models~\cite{Murase:2013rfa}. This suggests that the accompanying $\gamma$-rays produced via cosmic-ray interactions needs to be ``hidden'' by absorption in the neutrino sources~\cite{Murase:2015xka}.

The extragalactic $\gamma$-ray background observed by Fermi~\cite{Ackermann:2014usa} has contributions from  identified point-like sources on top of the IGRB shown in Fig.~\ref{fig1}. This IGRB is expected to consist mostly of emission from the same class of $\gamma$-ray sources that are individually below Fermi's point-source detection threshold (see, e.g., Ref.~\cite{DiMauro:2015tfa}). These underlying source populations can be deciphered via one-point fluctuation analyses~\cite{TheFermi-LAT:2015ykq,Lisanti:2016jub,DiMauro:2017ing} and the remaining isotropic $\gamma$-ray background allows to put stronger limits on neutrino emission models~\cite{Bechtol:2015uqb}. Alternatively, a significant contribution of $\gamma$-rays associated with IceCube's neutrino observation would have the implication that many extragalactic $\gamma$-ray sources are also neutrino emitters.

Another intriguing observation is that the high-energy neutrinos observed at IceCube could be related to the sources of UHE CRs. The simple argument is as follows: UHE CR sources can be embedded in environments that act as ``storage rooms'' for CRs with energies far below the ``ankle''. This energy-dependent trapping can be achieved via CR diffusion in magnetic fields. While these CRs are trapped, they can produce $\gamma$-rays and neutrinos via collisions with gas. If the conditions are right, this mechanism can be so efficient that the total energy stored in low-energy CRs is converted to that of $\gamma$-rays and neutrinos. The contribution of UHE CR sources to the diffuse neutrino flux is therefore bounded by a ``calorimetric limit''~\cite{Waxman:1998yy,Bahcall:1999yr}, which is indicated as green dashed line in Fig.~\ref{fig1}

Interestingly, the observed diffuse neutrino flux is just below this calorimetric limit. It is therefore feasible that UHE CRs and neutrinos observed with IceCube have a common origin. If this is the case, the neutrino spectrum should reflect the energy-dependent release of CRs from the calorimeters. Ultra-high energy CRs are strongly attenuated by resonant interactions with background photons, as first pointed out by Greisen, Zatsepin and Kuzmin~\cite{Greisen:1966jv,Zatsepin:1966jv} (GZK). This GZK mechanism is responsible for the suppression of the UHECR proton at the highest energies (green solid line) and predicts a detectable flux of cosmogenic neutrinos~\cite{Beresinsky:1969qj} (green dotted line). Future measurements of the diffuse PeV-EeV neutrino emission can provide supporting evidence for the UHE CR connection.

\section{Neutrino Emission from Blazars}

\begin{figure}[t]\centering
\includegraphics[width=0.95\linewidth]{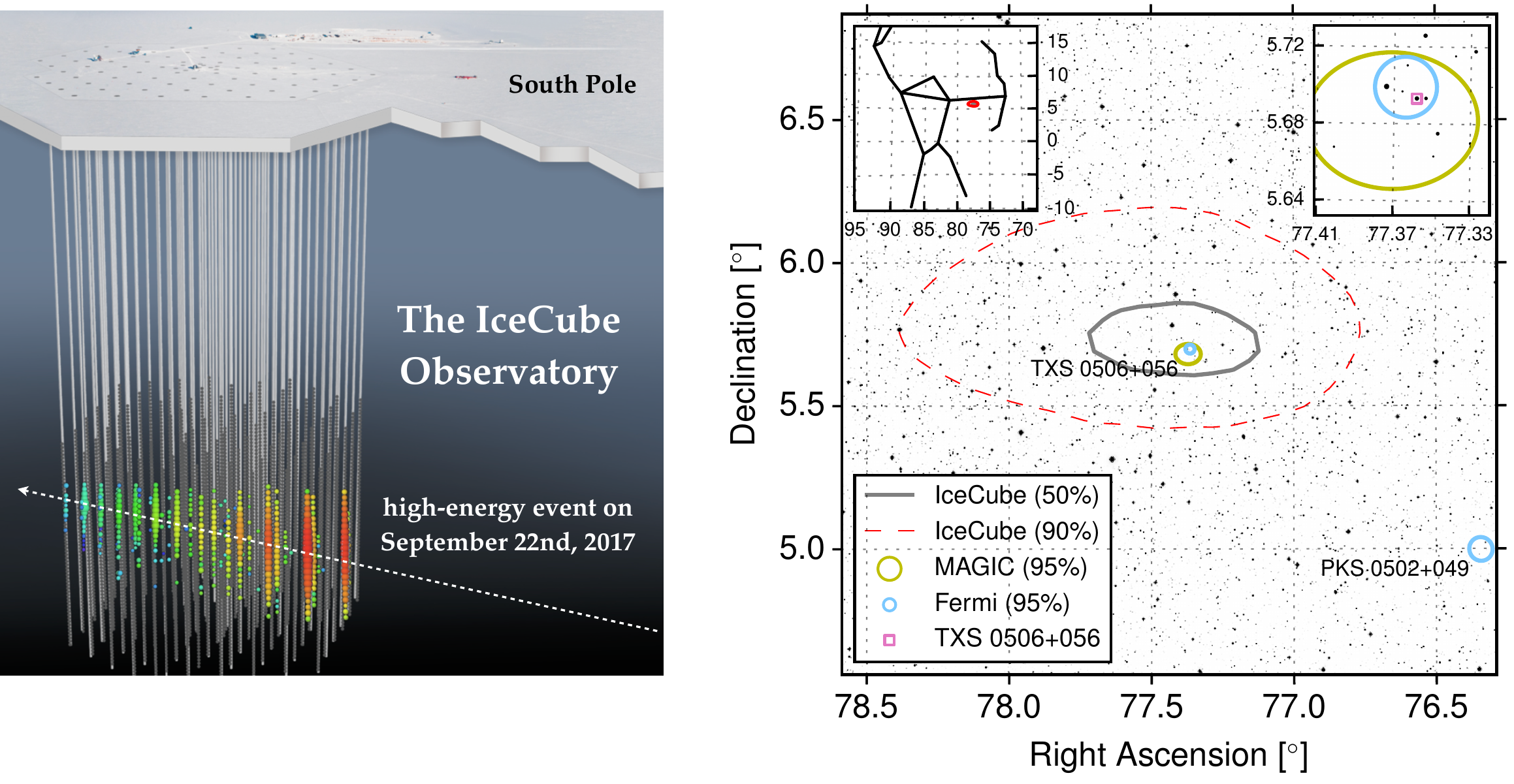}
\caption[]{{\bf Left panel:} A sketch of the IceCube Observatory at the South Pole. The detector consists of 5,160 digital optical modules (DOMs) that are attached along 86 support and read-out cables at a depth of 1,450 to 2,450 meters below surface. The alert IC-170922A is shown by colored DOMs, with color scale ranging from red to blue for early and late Cherenkov light detection, respectively. {\bf Right panel:} The blazar TXS 0506+056 located just off the left shoulder of the constellation Orion (top-left inset plot). The contours indicate the localization in $\gamma$-rays (Fermi \& MAGIC) and neutrinos (IceCube). (Figure from Ref.~\cite{IceCube:2018dnn}.)}\label{fig2}
\end{figure}

On September 22nd, 2017 IceCube observed a high-energy muon that entered the detector from a direction a few degrees below the Horizon (see left panel of Fig.~\ref{fig2}). These rare type of events trigger alerts that are sent out to partner observatories to find hints of electromagnetic emission from the same part of the Universe. After one week the Fermi satellite reported that the neutrino alert IC-170922A was spatially coincident with a known $\gamma$-ray source, the blazar TXS 0506+056, that was undergoing a period of enhanced $\gamma$-ray emission. The chance correlation of the neutrino alert with the $\gamma$-ray source is at the level of $3\sigma$ and provides evidence that the blazar was responsible for the event~\cite{IceCube:2018dnn}.

The blazar's spectral energy distribution at the time of the outburst can be well modeled by lepto-hadronic and proton-synchrotron models~\cite{Gao:2018mnu,Keivani:2018rnh,Zhang:2018xrr,Cerruti:2018tmc,Gokus:2018lgx,Sahakyan:2018voh}. It was pointed out that in these models predict a low neutrino flux ($\ll 1$~event) for the 2017 outburst that is limited by the observed level of the X-ray emission~\cite{Gao:2018mnu,Keivani:2018rnh} and the theoretically feasible proton luminosity, unless one considers enhanced pion production efficiencies in alternative models~\cite{Ahnen:2018mvi,Righi:2018xjr,Murase:2018iyl,Liu:2018utd}.

However, the low model predictions are not surprising. The emission of TXS 0506+056 has to be discussed in the larger context of the blazar population~\cite{Strotjohann:2018ufz}. The collection of blazar flares might individually only contribute with a low neutrino flux, but their sum can still provide observable event numbers. The question, which flare will eventually be responsible for the observed neutrino, becomes then simply a matter of chance. This is the essence of the {\it Eddington bias}, {\it i.e.} the trend to overestimate the intrinsic neutrino flux of a source observed with one neutrino: TXS 0506+056 was just the lucky winner of the cosmic lottery~\cite{Strotjohann:2018ufz}.

\begin{figure}[t]
\includegraphics[width=\linewidth]{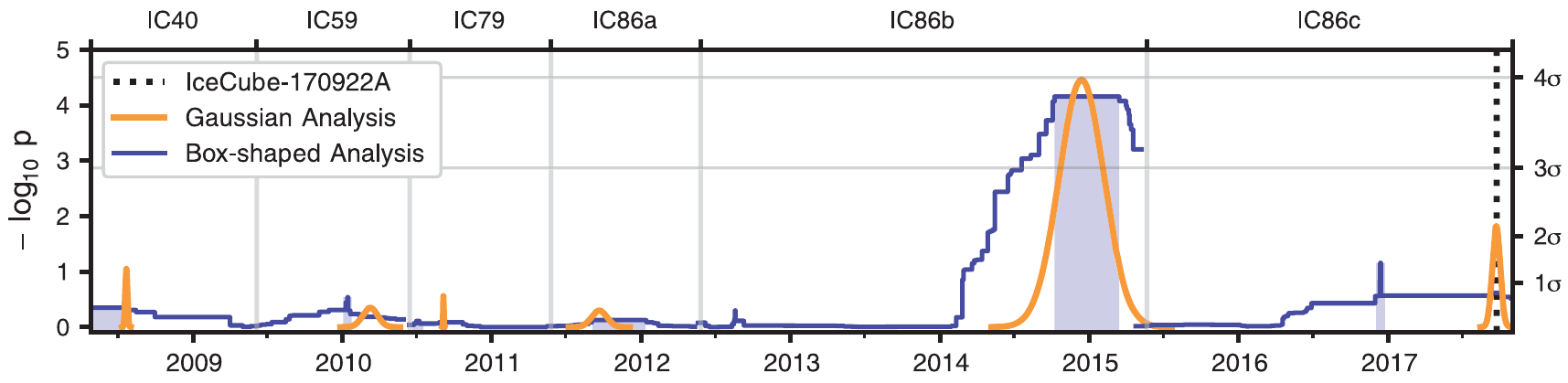}
\caption[]{The local $p$-values of past neutrino outbursts in the direction of TXS 0506+056. The orange and blue line shows results using a Gaussian and box-shaped neutrino time-profile. The vertical dotted line shows the time of the alert IC-170922. (Figures from Ref.~\cite{IceCube:2018cha}.)}\label{fig3}
\end{figure}

Motivated by the correlation seen by Fermi, IceCube investigated archival data for evidence of past neutrino emission of TXS 0506+056. Figure~\ref{fig3} shows the local $p$-values of past neutrino outbursts of this analysis using a Gaussian (orange) and box-shaped (blue) neutrino time profile. This analysis revealed that during a period from September 2014 to March 2015 the sources showed a prolonged outburst with an estimated $13\pm5$ neutrino candidates. A chance correlation of this type of neutrino outburst can be excluded at a confidence level of $3.5\sigma$~\cite{IceCube:2018cha}. Together with the earlier observation, this provides compelling evidence that this blazar is a source of high-energy neutrinos.

The 2014/15 outburst of neutrinos is not accompanied by a flare in the archival Fermi data. The corresponding neutrino luminosity is about four times larger than the $\gamma$-ray luminosity in the quiescent state. There is a statistically weak hint that the Fermi spectrum during the time of the neutrino outburst shows a hard spectral index, that could indicate an unobserved outburst of TeV $\gamma$-rays~\cite{Padovani:2018acg}. In any case, the different photon and neutrino emissions in the 2014/15 and 2017 observations, respectively, are presently the greatest challenge in finding a unified model of the neutrino emission mechanism.

Are blazars then also responsible for the diffuse TeV-PeV neutrino emission? This question was already addressed in an earlier study by IceCube~\cite{Aartsen:2016lir} looking for the combined time-integrated emission of blazars observed in $\gamma$-rays. This study could not find evidence for neutrino emission and placed an upper limit on the relative contribution of blazars to the diffuse neutrino flux. For an $E^{-2.5}$ spectrum this upper limit is at the level of 27\%. In light of the recent results, it is therefore plausible that blazars have a large contribution, but it is unlikely that they dominate the observed diffuse neutrino flux.

\section{Conclusions}

The future of neutrino astronomy is promising. Recent observations with IceCube have revealed a diffuse TeV-PeV neutrino flux of unknown origin. The intensity of the neutrino flux is comparable to that of $\gamma$-rays and ultra-high energy cosmic rays, which provides excellent conditions for multi-messenger studies. We now also have first compelling evidence that blazars are sources of high-energy neutrinos, but they are unlikely the sole contributors of the diffuse flux. With the help of future optical Cherenkov telescopes -- KM3NeT~\cite{Adrian-Martinez:2016fdl}, Baikal-GVD~\cite{Avrorin:2018ijk}, and IceCube-Gen2~\cite{Aartsen:2014njl} -- we will be able to shed more light on  these recent observations and establish high-energy neutrinos as an essential component for multi-messenger astronomy.

\section*{Acknowledgements}
I would like to thank the organizers of the {\it 7th Roma International Conference on Astroparticle Physics} for their invitation to present this work. I would also like to thank my colleagues in the IceCube collaboration for their support. This work was supported by Danmarks Grundforskningsfond (project no.~1041811001) and \textsc{Villum Fonden} (project no.~18994).

\bibliography{biblio.bib}

\end{document}